\documentclass[english,gc, manuscript]{copernicus}
\usepackage[T1]{fontenc}
\usepackage{array}
\usepackage{longtable}
\usepackage{rotfloat}
\usepackage{multirow}
\usepackage{pdflscape}
\usepackage{graphicx}
\usepackage[authoryear]{natbib}

\makeatletter

\providecommand{\tabularnewline}{\\}

\makeatother

\usepackage{babel}
\begin{document}
\nolinenumbers
\title{SSFX (Space Sound Effects) Short Film Festival: Using the film festival
model to inspire creative art-science and reach new audiences}
\author[1]{Martin O. Archer}
\affil[1]{School of Physics and Astronomy, Queen Mary University of London,
London, UK}
\correspondence{Martin O. Archer\\
(m.archer@qmul.ac.uk)}
\runningtitle{SSFX (Space Sound Effects)}
\runningauthor{Martin O. Archer}
\maketitle
\begin{abstract}
The ultralow frequency analogues of sound waves in Earth\textquoteright s
magnetosphere play a crucial role in space weather, however, the public
is largely unaware of this risk to our everyday lives and technology.
As a way of potentially reaching new audiences, SSFX made 8 years
of satellite wave recordings audible to the human ear with the aim
of using it to create art. Partnerting with film industry professionals,
the standard processes of international film festivals were adopted
by the project in order to challenge independent filmmakers to incorporate
these sounds into short films in creative ways. Seven films covering
a wide array of topics/genres (despite coming from the same sounds)
were selected for screening at a special film festival out of 22 submissions.
The works have subsequently been shown at numerous established film
festivals and screenings internationally. These events have attracted
diverse non-science audiences resulting in several unanticipated impacts
upon them, thereby demonstrating how working with the art world can
open up dialogues with both artists and audiences who would not ordinarily
engage with science.
\end{abstract}

\introduction{}

Public engagement projects which see artists and scientists collaborate
together in some way have become increasingly popular, with growing
evidence that such projects, through a variety of methods, contribute
to the development of society \citep{malina18}. \citet{malina10}
classifies such collaborations into the following categories:\renewcommand{\labelenumi}{\Roman{enumi}.}
\begin{enumerate}
\item Scientists collaborating with artists on common projects resulting
in both scientific discoveries and the production of art works
\item Scientists applying scientific research to understand creative activity
in the arts
\item Scientists working with artists to develop technological inventions
\item Working as both a scientist and an artist in a dual career
\item Scientists engaging with the arts to enhance cultural appropriation
of science
\item Scientists engaging with the arts to improve the ways science is communicated
to the public
\end{enumerate}
\renewcommand{\labelenumi}{\arabic{enumi}.}To better understand scientists'
motivations for such endeavours at an art-science session at the 2019
Interact symposium \citep{interactsession19} 12 university science
researchers and public engagement professionals were surveyed for
their attitudes towards art-science collaborations. This found that
their interest in art-science collaborations were based on enjoyment
of both subjects, utilising their creativity, and as a communication
tool (particularly for different audiences). When asked in an open
question who they thought the audience was for such collaborations:
$67\pm17\%$ thought they are for everyone; $25\pm16\%$ said non-science
arts audiences; and one person responded it depended on the aims of
the project (see Appendix~\ref{sec:Statistical-techniques} for details
of statistical techniques used throughout and Appendix~\ref{sec:Qualitative-data}
for all the responses). Respondents thought art-science is important
because it provides different ways of communicating science, can reach
new audiences, can help embed science as part of culture, and that
both disciplines can learn from one another through their respective
creativity. Therefore Type VI of \citet{malina10} was the most highly
cited typology of art-science collaboration, though Types IV and V
were also mentioned.

There have been numerous published examples of science-inspired artworks
(Type V), where science acts as a resource for creative art \citep{kim11}.
\citet{vossandreae} presents sculptures inspired by quantum physics
that he argues can indicate aspects of reality that science cannot.
The Tumamoc Hill Arts Initiative was a collection of site-based art
and writing inspired by the Sonoran Desert and the underlying science
of the region \citep{mirocha15}. Similarly, \citet{orfescu12} describes
artistic interpretations of scientific images, in this instance nanostructures,
where artists convert them into pieces of art. \citet{hoare13} posits
that even classic works of literature, such as `Moby Dick', have strong
scientific influences since art and science were not strictly demarcated
at the time. It is therefore clear, even from these few examples,
that activities attempting to integrate science into culture are incredibly
varied and have been undertaken for a long time.

Engaging with new audiences seems to be a prominent motivation for
scientists in undertaking art-science collaborations and many evaluations
of art-science events have tried to assess whether they have indeed
attracted non-science audiences. Science et Cit\'{e}, a festival
across 20 cities in Switzerland, while striving to be \textquotedblleft a
festival of the sciences and arts'' attracted significantly more
people who were interested by science ($40\pm1\%$) than art ($24\pm1\%$)
(respondents could select from any number of 14 options), with the
festival's more art-themed events typically only attracting $1.4\pm0.3$
times more people with arts interests than science ones \citep{vonRoten07}.
Another example --- Covariance, a month-long art-science exhibit
in London --- found $95\pm1\%$ of their audience were frequent or
occasional art goers and $83\pm3\%$ attended science events, hence
there was substantial overlap in these two areas ($\geq78\pm3\%$)
\citep{lynch13}. Finally, the Art and Space exhibition in Dunedin,
New Zealand attracted audiences $57\pm10\%$ of which had a professional
background in the arts compared to $26\pm9\%$ in science, who primarily
attended due to a general interest in art ($71\pm9\%$) rather than
science ($38\pm9\%$) though $50\pm10\%$ were attracted by how science
and art can combine \citep{brook17}. These case studies therefore
highlight that art-science events vary significantly in their audiences'
interests and do not necessarily always attract new audiences as desired.

Given the multitude of different formats that constitute the art world,
there are many ways of combining it with science. The twentieth century
saw film emerge as one of the main art forms readily appreciated by
the public \citep{nowellsmith17} and in recent years the film festival
has burgeoned into an important area of cinema, both culturally and
industrially, with an incredibly diverse range of festivals running
internationally \citep{archibald11}. Research into film festival
attendees \citep{baez14} has revealed three key motivating factors:
``discovery'', ``entertainment'', and ``cinema''; with specialised
film festivals (such as those covering specific genres, topics or
issues) also providing a general feeling of belonging to a specific
group and/or ``cinephile community'' \citep{devalck09,ffrn}. Film
festivals surrounding science have been growing in number \citep[e.g.][]{eurasf,imagine}
with these typically featuring documentary films presenting scientific
findings in an entertaining but still educational way. However, beyond
simply improving the ways that science is communicated to the public
(Type VI of \citet{malina10}), there is the potential to as well
have science appreciated more as part of culture via film (Type V).
BIO-FICTION invites short films addressing current/future debate topics
in areas of biology, with a near even mix of fiction and documentary
style submissions \citep{schmidt15}. CineGlobe is a film festival
at CERN which centres around broad and culturally relevant themes
inspired by science and technology \citep{cineglobe}. CineSpace is
a film festival by NASA and Houston Cinema Arts Society which solicits
films inspired by and using actual NASA imagery.

This paper concerns a film festival project called SSFX (Space Sound
Effects), devised and run by the author, which aimed to integrate
space science research into culture. The scientific basis for the
project was the ultra-low frequency (ULF) analogues of sound present
within near-Earth space \citep[e.g.][and references therein]{Keiling2016}
which had been converted into audible sound \citep{archer18}. The
motivations for choosing to use these sounds for the creation of art,
and in particular through film, are discussed in section~\ref{sec:Motivations}.
The SSFX project had two phases, both with different target audiences
and aims. Phase one targeted filmmakers, aiming to engage the independent
filmmaking community with the sounds present in the near-Earth space
environment and enable the creation of creative short films inspired
by and incorporating these sounds. This was tackled by running an
international short film competition (adopting standard film festival
practises through partnering with film industry professionals) which
challenged filmmakers to use the sounds as key creative elements.
Section~\ref{sec:filmmakers} concerns this phase of the project
and the subsequent collaborative relationships that formed between
scientists and filmmakers through the project. It was through these
relationships that phase two of SSFX was possible, which aimed to
exhibit these films to wide and diverse audiences, exposing them to
this area of space science research and hence positively impacting
upon these non-traditional audiences. This phase therefore had two
target groups, film exhibitors/programmers and independent film-goers.
Section~\ref{sec:events} discusses how film exhibitors and programmers
were engaged to integrate the films into their events and venues,
whereas section~\ref{sec:audiences} concerns evaluating the backgrounds
of the audiences that attended these events and what impacts resulted
from them.

\section{Motivations\label{sec:Motivations}}

Space is far from completely empty, it's pervaded with very tenuous
plasmas such as the solar wind that streams off of the Sun. Earth's
intrinsic magnetic field acts as an obstacle to this wind and results
in a magnetosphere, protecting us from much of this harmful ionising
radiation. However, the interaction between the solar wind and the
magnetosphere is highly complex and dynamic, resulting in phenomena
which affect the space- and ground-based technology we increasingly
rely upon in modern life such as electrical grids, GPS systems, and
weather forecasts. These effects are known as space weather and have
been recognised as a potential risk to our everyday lives \citep{cannon13},
however, a large fraction of the general population are not aware
of this \citep{spaceweatherdialogue15}.

One way in which solar wind energy and momentum are transferred into
and around magnetospheres are through plasma waves. The spatial and
time scales where the weak plasma can largely be treated as a single
conducting fluid dictated by magnetohydrodynamics necessitates plasma
waves to fall within the ultra-low frequency (ULF) regime, with frequencies
fractions of milliHertz up to 1~Hz. Of course this does not lie within
the human auditory range, however, simply by dramatically speeding
up playback of satellite observations it is possible to make our ULF
wave measurements audible. \citet{archer18} converted perturbations
in magnetic field data (which move similarly to the plasma itself
due to the frozen-in condition within magnetohydrodynamics) taken
at geostationary orbit into an audio dataset which is now publicly
available from the National Oceanic and Atmospheric Administration
\citep{noaa}. This has already been used as tool in exploratory citizen
science with schools, as detailed by the authors, but could lend itself
to artistic uses also.

\begin{figure}
\begin{centering}
\includegraphics[width=0.5\columnwidth]{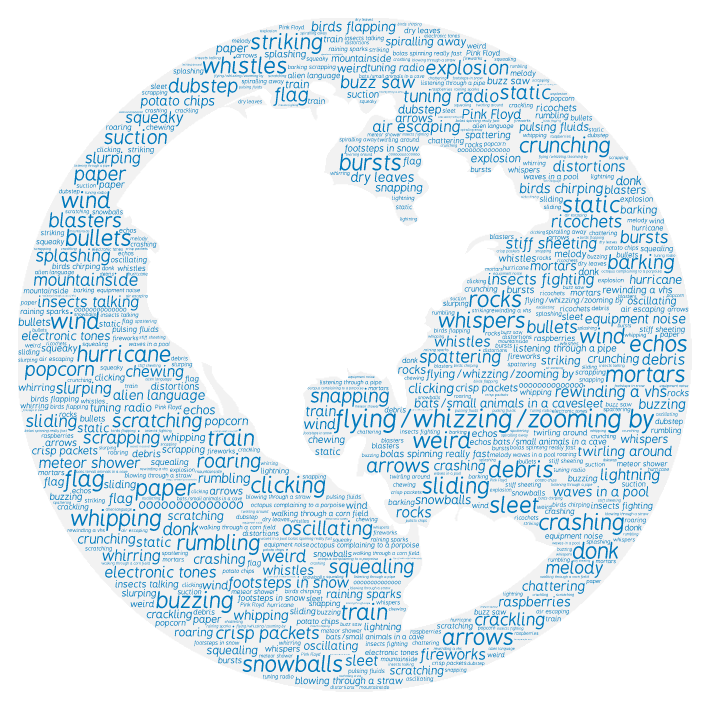}
\par\end{centering}
\caption{Word cloud of comments on what space ``sounds like'' from SoundCloud
playlist of space sounds.\label{fig:wordcloud}}

\end{figure}

How this audible version of scientific data, or ``sounds of space'',
could potentially be used in the creation of art was primarily informed
by the sounds themselves. They surprisingly did not typically take
on the musical quality somewhat expected by the researchers who study
discrete frequencies and resonances within Earth's magnetosphere,
but instead conveyed a sense of dynamism and variety as well as having
a somewhat cacophonous nature. While the audible dataset described
in \citet{archer18} was comprehensive enough in order to undertake
science, for the purpose of creating art there was redundancy. Therefore
to reduce the amount of data, only the time-differenced stereo summary
files were used, averaging these over all spacecraft to result in
only one audio file per year. Through an article in The Conversation
\citep{archerconversation16}, republished by Newsweek, Daily Mail,
Space.com amongst others, online comments using SoundCloud on what
people thought random periods of the data ``sounded like'' were
solicited. While the accuracy and precision of SoundCloud comments'
time-tagging meant it could not be used as a means of event identification
for scientific research, it did result in the wide range of 85 unique
responses (from 151 comments) that are shown in Figure~\ref{fig:wordcloud}.
These reflections on the sounds planted the idea of their usage as
sound effects in films, ultimately resulting in the SSFX (Space Sound
Effects) Short Film Festival project. In the following sections we
detail the various phases and audiences engaged through the project
from filmmakers and exhibitors to film-goers, presenting findings
on their motivations for getting involved and what impact the project
had on them.

\section{Inspiring creative art-science films\label{sec:filmmakers}}

\subsection{Establishing a film competition}

It was clear that even with the idea of having the sounds used in
creative ways within films, much expertise and advice from within
the film industry was required. We were interested in engaging with
the filmmaker community and seeing what they came up with themselves,
rather than commissioning something specific from a given filmmaker,
therefore we went down the public call model. We solicited expert
external contacts from Queen Mary's film department and as well as
existing film contacts. These individuals consulted us on how film
festivals operate and are run as well as pointing us in the direction
of several organisations and networks that would be helpful. It was
deemed that adopting standard film festival practices and establishing
film industry partnerships were vital in order to make the project
as authentic and attractive as possible for independent filmmakers,
a concern given this was a new initiative being spearheaded by scientists.
However, the film industry professionals we approached to voluntarily
sit on the judging panel found the concept exciting:\begin{quote}
\textit{``I found the project to be innovative, having never worked
with a shorts programme dedicated to engaging filmmakers with pre-recorded
sounds, space sounds, or an academic programme}'' (film industry
judge 1)\end{quote}

\begin{quote}\textit{``Many people forget that sound is one of the
most important aspects of good filmmaking\textquotedblright{}} (film
industry judge 2)\end{quote}They were joined on the panel by a couple
of scientists with experience in art-science collaborations.

There were necessarily some differences with the SSFX (Space Sound
Effects) Short Film Festival to most film festivals. Typically these
opportunities allow filmmakers to submit existing works with only
a few limiting criteria such as genre. However, we were challenging
filmmakers to incorporate very specific elements, the provided sounds
of space, into their work and in many cases making a film especially
for the festival. Given these unusual constraints, it was decided
that we would try to make the rest of the competition's criteria as
broad and inclusive as possible. Therefore we would not charge a submission
fee, there would be no restrictions on genre or topic, we would allow
filmmakers to modify the sounds as they saw fit, and permit films
created specifically for the competition or existing films edited
to integrate the space sounds. The only other criteria we set were
by age and location, with categories initially for both UK and international
filmmakers separately in the age ranges: under 18, 18--24, and 25+.
The significant work involved on the filmmaker side necessitated there
being a large submission window, which we set as six months long,
which would hopefully provide enough time to produce high quality
short films. We felt it was important for there to also be monetary
prizes associated with the competition to ensure that filmmakers\textquoteright{}
efforts were valued.

A website was established which hosted the space sounds for download,
more information about the competition/festival, and would post YouTube
videos throughout the submission window providing more background
on the science \citep{ssfx}. However, it was deemed that using an
existing online film festival submission platform would be better
than coming up with our own method. Desk research highlighted two
portals --- Withoutabox and \citet{filmfreeway}. We opted for the
former, given it was the first online film festival submission service
and was owned by IMDB. In hindsight, however, we realised that Film
Freeway would have been more flexible. Withoutabox subsequently closed
down in late 2019. While we had set our final submission deadline,
staff at Withoutabox recommended within their system that we have
various different stages of deadlines (``early bird'', ``standard''
etc.) since this would flag the opportunity to filmmakers looking
at Withoutabox's upcoming deadlines calendar. Finally, in order to
reduce ineligible entries we asked that filmmakers provide some information
on how they used the space sounds. At first this was simply written
in the terms and conditions to be included in their cover letter.
However, it soon became clear that many filmmakers were not reading
the terms and simply submitting their ineligible films anyway. We
were able to get Withoutabox to add a custom required question which
explicity asked the filmmakers to provide this information, which
dramatically cut down (but did not entirely eliminate) spam entries.
At this stage the competition was open and we simply needed filmmakers
to engage with the opportunity.

\subsection{Engaging with filmmakers}

To share the opportunity widely within the independent film community,
existing networks were utilised: a protracted marketing campaign throughout
the submission window to Shooting People's over 45,000 member base
was run through newsletters, an editorial feature, and social media
\citep{shootingpeople}; flyers about the competition were mailed
to every film school in the country; we attended London-based filmmaker
Meetup groups discussing the opportunity with around 70 filmmakers
\citep{meetup}; and we contacted key people recommended by film industry
judges for more grassroots marketing. As part of formative evaluation
to ensure these were being effective at reaching our target audience,
we monitored the number of people who registered interest in the project
on our website (essentially subscribing to a mailing list) recording
also their age, location, and what their motivations for signing up
were. In total 102 people signed up, after having discarded spam entries
(see later). The majority of people were 25 or over at $62\pm6\%$,
with few under 18s at only $10\pm3\%$ (in hindsight perhaps to be
expected), which informed our merging of the two younger age categories
in the competition early in the submission window. In terms of location
there was an almost even split in absolute terms between those from
London, elsewhere in the UK, and internationally, which is clearly
unrepresentative of the global population and likely down to the main
networks used to promote the opportunity.

\begin{figure}
\begin{centering}
\includegraphics{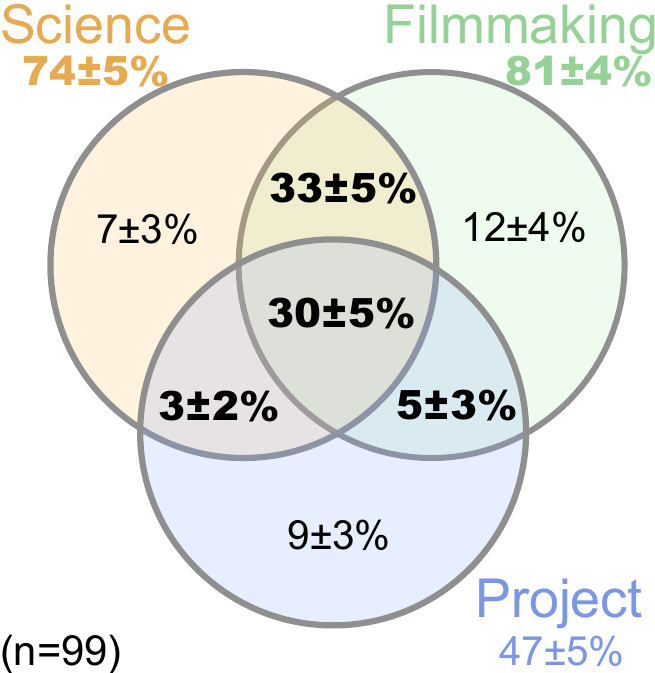}
\par\end{centering}
\caption{Venn diagram of people's reasons for registering interest with SSFX.
Bold values denote statistically significant differences from pure
randomness (expected 57\% in each overall set, 14\% in each region)
taking into account multiple comparisons ($\alpha_{Bonf}=0.017$ for
sets and $\alpha_{Bonf}=0.0071$ for regions). Created in part using
InteractiVenn \citep{heberle15}.\label{fig:filmmaker-venn}}
\end{figure}

To assess people's motivations we asked them to select from as many
of the following options as applied: an existing interest in science
generally ($S$), an existing interest in filmmaking ($F$), interest
specifically in the project ($P$), or some other reason which they
could then specify. We assume that all entries which did not select
any of these options (including other) were spam. Sixteen people reported
other reasons for registering interest (with three not selecting any
of the main options): nine had a background in either sound design
or musical composition; three were considering visualising the sounds;
with others mentioning the creative challenge, interests in space
or art-science, and the possibility of integrating the resulting films
into existing science or art-science events. Figure~\ref{fig:filmmaker-venn}
shows the breakdown over the three main options. The proportions in
each set (i.e. $S$, $F$, and $P$) and region (of the Venn diagram)
have been compared to those expected purely at random, with $S$,
$F$, and $S\cap F$ being significantly greater than expected. We
also compared the sizes of all sets and unions of sets with one another
finding that most of these differences are statistically significant
--- of the 6 possible comparisons only $S$ vs. $F$ and $F$ vs.
$S\cup P$ were not. From this we deduce that people who registered
interest typically had existing interests in both science and filmmaking.
Given only small (typically only a few percent) fractions of the public
work or have qualifications in film \citep[e.g.][]{bfi19}, we conclude
that SSFX successfully engaged the filmmaking community. Given the
considerable effort involved in creating a film, an interest in science
also is thus understandable, though anecdotally from conversations
with filmmakers it was found that their primary scientific interests
(if any) were typically not in physics or space science though. 

Leading up to the competition's deadline very few films had been submitted,
therefore to assess this we sent out a survey to the mailing list
six weeks before the deadline. While only seven responses were received,
six indicated they intended to submit films though only two of these
were confident that their film would be ready by the deadline. Based
on these results we decided to extend the deadline by an additional
two weeks to allow more time for filmmakers, while still having sufficient
time for judging and event organisation. Even closer to the submission,
a few filmmakers reached out stating that their films would not be
complete so we took the decision to allow work in progress submissions,
so long as the filmmakers indicated what additional work needed doing
and that it could be achieved in time for the event.

\subsection{Evaluating film submissions\label{subsec:submissions} }

By the deadline 22 eligible films had been submitted (180 ineligible
films not featuring any space sounds were also submitted, most of
which came before the bespoke question was implemented), which according
to their credits involved a total of 90 people. These films themselves
demonstrate an impact on filmmakers, given that they have engaged
with an opportunity to co-create an art work based around scientific
data --- a substantial undertaking. Most entries (nine) were in the
25+ UK category with 4--5 entries in the other three categories.
This also meant that most entries came from the UK ($59\pm12\%$)
though we note international filmmakers from Brazil, Canada, Italy,
Portugal, and USA also submitted films. None of the differences in
submission numbers were statistically significant by category, age,
or internationality.

Each film was scored by the judges on both their usage of the space
sounds (e.g. a few submissions had just a token usage of the sounds
within their films) and overall impression of the film with equal
weighting within Withoutabox's online system. Judges could also leave
any written feedback on both judging criteria to help final decisions.
A subset of all the submissions based on total runtime were assigned
to each judge, though 11 films were seen by all judges (the shortest
ones) and at least 3 judges saw each submission. One of the film industry
judges noted:\begin{quote}\textit{``The process of running the competition
was extremely professional and I would recommend the model to others
in the future, with a secure screening system, showing full creative
credits for each film that allowed feedback to be added and votes
cast within one dedicated site. Thanks for such great organisation
and clear steer on how you wanted the judging to go.''} (film industry
judge 1)\end{quote}In the end there was a fair amount of disagreement
between the judges --- the alpha coefficient of \citet{krippendorff70,krippendorff}
for these ordinal measurements was only $0.43$ (where a value of
1 would indicate perfect agreement and 0 would result from randomly
drawn scores, see Appendix~\ref{sec:Statistical-techniques}). Each
judge's scores were therefore standardised, using means and standard
deviations across only those films which were seen by all, to ensure
no one judge had more sway in the outcome. Given the overall time
within the venue we were able to select the top eight films (based
on the average standardised scores) for exhibition, however, one of
these was unfinished upon submission and could not ultimately be completed
in time. This film dropping out necessitated merging the two international
categories.

One of the film industry judges noted about the submissions\begin{quote}\textit{\textquotedblleft I
was really impressed by the quality and diversity of films submitted
through the competition as well as the international uptake. The range
of film making styles was really interesting, there were dramas, comedies,
animation, science fiction and avant-garde productions with some films
exploring the scientific concepts directly and others using them in
more abstract ways... I really loved the fact that the project was
open in how filmmakers could interpret the sounds in their productions
and I think this was key to gaining the variety that appeared across
the submissions.''} (film industry judge 2)\end{quote}We note that
even while some of the films incorporated elements of the underlying
scientific concepts, they were not the primary focus of the films
and were done in creative fictional ways and not in a documentary
style. Thus all the films fell under the ``science as culture''
type of art-science, i.e. Type V of \citet{malina10}, rather than
a form of direct science communication (Type VI). Even one of the
judges noted ``\textit{it has genuinely got me thinking about how
I could explore some of the research in my future creative outputs''}
(film industry judge 2). The conclusion from these is that by giving
a huge amount of creative freedom over to the filmmakers in allowing
them to interpret and include this scientific data as they saw fit,
it enabled the creativity and variety of films submitted, thus highly
aligning with the aim of integrating science into culture. The following
sections summarise the selected works, including the filmmakers' reflections
captured during filmed panel discussions at events.

\subsubsection{Astroturf}

Synopsis: A meticulous young man tends to his fake garden to the sounds
of deep space\\
Genre: Science Fiction\\
Duration: 1~min

The film depicts a man performing gardening tasks, though this garden
features no real plants instead being filled with the titular astroturf
along with plastic flowers, trees, butterflies etc. It is revealed
at the end that this garden in on the moon and that the Earth is on
fire. The director noted\begin{quote}\textit{``We wanted to make
a film that used the space sound effects in an interesting way, while
telling a compelling short science-fiction story. The rustling, swirling
space sounds reminded us of the noises that people make all the time
when performing simple tasks - sounds that in film are often replaced
or reproduced as foley. So we decided to build the entire soundtrack
from the space sound effects, and created a simple narrative that
involved a combination of actions that we felt would be convincing
when dubbed. We came up with the idea of putting it in space because
of the space sounds... Because {[}in the film{]} we've screwed up
the Earth, for him this tiny patch of land is extraordinarily precious
and so that was where his character emerged from.'' }(director of
`Astroturf')\end{quote}The producer added\begin{quote}\textit{``He's
trying to recreate what he's known in the past, what a real garden
is, in this fake world that he's living it. Because we're both 'greenies'
and we see that as a potential future for us, that was what really
inspired that incongruess nature of nature versus fakeness.'' }(producer
of `Astroturf')\end{quote}Further comments can be found at https://www.youtube.com/watch?v=RpC-sFzUnEE.

\subsubsection{Dark Matter(s)}

Synopsis: An experimental and meditative imagining capturing the activities
of a fish tank in a way that takes the inhabitants out of their enclosed
world, to a place unknown, to feel both their death and their life.\\
Genre: Video Art\\
Duration: 5~min

The director describes the film as\begin{quote}\textit{``about a
couple fish in a fish tank, but we tried to film it in a way that
it doesn't look like they're in a fish tank... that got rid of the
boxed in enclosure. When I realised what's bigger than being a locked
in a box was everything, it made such sense to me to look into sounds
from space. I think the sound from space gives it that extra push
for {[}the fish{]} to like break out of this cage or for that dichotomy
of inside versus outside to be transcendent. As soon as I figured
out that I was going to use sound effects from space I think the project
came full circle.''} (director of `Dark Matter(s)')\end{quote}The
sound designer commented on how he modified the sounds for the film,
which were matched with classical music throughout:\begin{quote}\textit{``I
used a lot of reverbs to soften them {[}the sounds{]}, I did a little
bit of slow panning, I shifted the pitch on the sound effects a couple
times to separate them, I shifted them down to make them a little
deeper. There's this one part where there's a bunch of bubbles and
so I changed them so they could sound like bubbles. Basically the
whole process was making them soft enough to fit in the context of
the film.''} (sound designer of `Dark Matter(s)')\end{quote}Further
comments can be found at https://www.youtube.com/watch?v=quLaFmS9kDE.

\subsubsection{Murmurs of a Macrocosm}

Synopsis: A journey through a microscopic world. We are led via the
descriptive recordings of those who travelled it.\\
Genre: Science Fiction\\
Duration: 5~min

This film shows recoloured drone footage from Snowdonia paired with
NASA Apollo recordings and the space sounds to depict exploration
in a microscopic realm, which is revealed at the end to be inside
the grooves of a vinyl record of the moon landings. The filmmaker
stated\begin{quote}\textit{``It was a visual that I always loved
to, those SEM microscopic images that are colourised black and white
images. But I always wanted a little bit more, I wanted to move around
them. I think when hearing those sounds it kind of reminded me a bit
of a record player. It also reminded me a lot of the sound from space
in 'Contact', the Carl Sagan film / book, that's how it came together
from those little things.''} (director of `Murmurs of a Macrocosm')\end{quote}When
discussing the use of NASA recordings he noted\begin{quote} \textit{``finding
this sort of innate humour and human conversation that they would
often have to each other and they just felt so in awe the whole time...
they're constantly excited which I really I love that aspect of exploration.}''
(director of `Murmurs of a Macrocosm')\end{quote}Further comments
can be found at https://www.youtube.com/watch?v=e3kUGlvI\_Hk.

\subsubsection{Names and Numbers}

Synopsis: A sound and voice collage shaped by the sounds of space
and Morse code, addressing the external, physical and material experiences
of sound and movement contrasted with interior reflections, explored
through language, inner voices and symbols.\\
Genre: Experimental\\
Duration: 14~min

The filmmaker explained how this experimental piece came together:\begin{quote}\textit{``I
tried to enact that experiment of writing down your thoughts to the
sound of a buzzer which samples your mind at any particular time...
It was basically an accumulation of ideas and just sitting down and
following the logic of each individual material thing: a soundtrack,
recording a piece of text, a collection of different images. There
was no simple way of putting them all together and I guess the stream
of consciousness of that writing process was one of the guiding principles.}''
(director of `Names and Numbers')\end{quote}Further comments can
be found at https://www.youtube.com/watch?v=Uuvcm1YfdZ4.

\subsubsection{Noise}

Synopsis: A secretive woman opens herself up to her unruly housemate,
after they are stuck together in her room.\\
Genre: Drama\\
Duration: 13~min

This drama film is about a woman who often isolates herself by listening
to the sounds of space, and who doesn't get along with her very different
housemate. They eventually are able to connect over these sounds.
The director noted on the title\begin{quote}\textit{``noise is a
specific scientific term for something which has no informational
value... and so when the characters are talking to each other they're
trying to work out what's noise and whether they can actually understand
anything from what they're saying to each other... Once you heard
the sounds they kind of wrote the story, they had to carry the narrative,
creating a character in and of themselves.'' }(director of `Noise')\end{quote}Further
comments can be found at https://www.youtube.com/watch?v=Fgvo\_lP7ZmA.

\subsubsection{Saturation}

Synopsis: There's no answer when time is the question. Featuring 35mm
slides found in a medical archive, this sci-fi story concerns unknown
phenomena that made all organic processes so fast as to make life
impossible.\\
Genre: Science Fiction\\
Duration: 7~min

This film couples still images of medical imagery with subtitled text
and a soundtrack composed of modified space sounds. The filmmaker
explained the creative process\begin{quote}\textit{``When I first
started to edit the film a couple of years ago, before I even knew
what it was supposed to be, I thought {[}the mysterious phenomenon
in the film's narrative{]} was something related to space... When
I saw your call {[}for films using the sounds of space{]}, I realised
that's what I needed - something really from space that I can use
on my film. To make the sounds more tense I saturated them, making
them more drone-like... I was also interested by the process itself
of making the sounds hearable by stretching and compressing the time
and it is very related to the narrative that I was thinking.}'' (director
of `Saturation')\end{quote}Further comments can be found at https://www.youtube.com/watch?v=rYxFHExQ4aQ.

\subsubsection{The Rebound Effect}

Synopsis: Bringing together contemporary movement and digital media
to capture dance in a way which pushes beyond the tangible dimensions
of live performance.\\
Genre: Dance / Music Video\\
Duration: 2~min

This film depicts a modern dancer moving to the sounds from space
mixed with electronic music. Unfortunately filmmaker comments about
this work were not captured.

\subsection{Running a film festival}

A boutique arts and cinema venue was hired for the SSFX Short Film
Festival (Rich Mix in Shoreditch, London). To capitalise on their
regular members, we opted to have them host ticket sales and primarily
undertake marketing for the event. While the event was not being run
to make a profit, we decided to charge a small ticket price to reduce
cancellations and convey a sense of perceived value for the event.
In reality all ticket proceeds actually went into the cost of a free
post-event reception. The cinema required the films in a digital cinema
package (DCP) format. First we received the high-quality video files
from the selected filmmakers and then converted these using the free
open-source \citet{dcpomatic} software. For exhibition to the public
the films required British Board of Film Classification (BBFC) certification
also, which were submitted as DCPs online. The majority of classifications
were Universal rating bar two --- one was deemed Parental Guidance
for ``mild surgical detail'' and another gained at 15 rating due
to ``strong language and drug misuse''.

\begin{figure*}
\begin{centering}
\includegraphics[width=0.75\columnwidth]{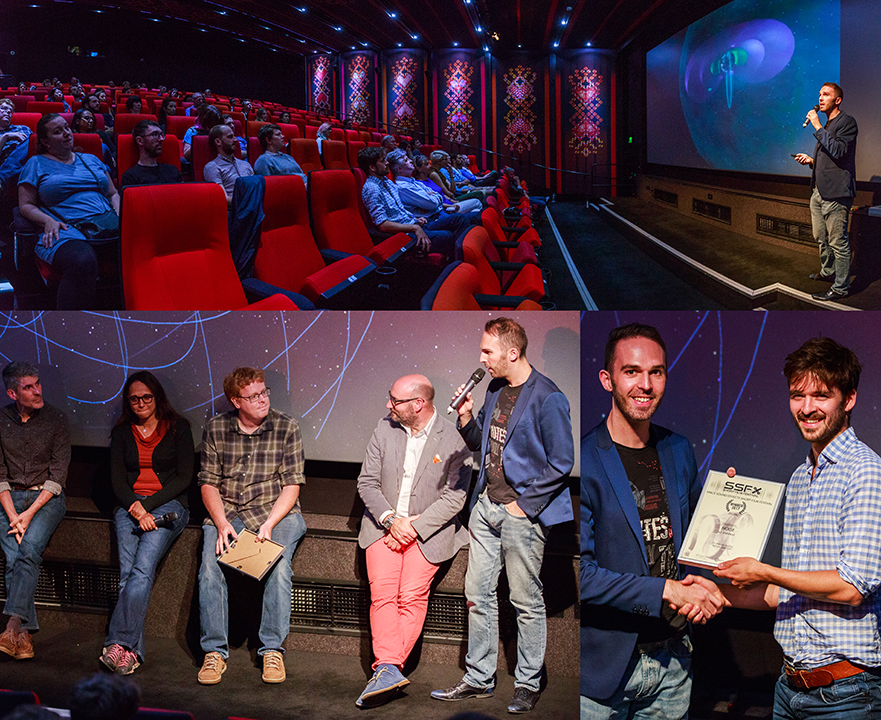}
\par\end{centering}
\caption{Photos from the SSFX Short Film Festival. (Top) Presentation about
the underlying science. (Bottom Left) Panel discussions between scientists
and filmmakers. (Bottom Right) Awarding of prizes to filmmakers.\label{fig:Photos}}

\end{figure*}

The event was started with an unadvertised short presentation on the
underlying science to the audience. These were then followed by groups
of film screenings, awards presentations, and panel discussions between
scientists and filmmakers (international filmmakers joined via video
conferencing) about their work and approach to the project. Photos
of all of these can be seen in Figure~\ref{fig:Photos}. The post-event
reception then enabled further discussion between scientists, filmmakers,
and film-goers. Evaluation of this event (and subsequent ones) can
be found in sections~\ref{sec:events} and \ref{sec:audiences}.

\subsection{Supporting the films and filmmakers}

Following the SSFX Short Film Festival, we wanted to support the filmmakers
in sharing their work more widely. In return we asked them to add
specific prologue/epilogue text about the underlying science as well
as items in the credits pertaining to project staff, data providers,
and funders. In hindsight, it may have been easier to ask for this
at the selection stage so these would have been incorporated into
the high quality versions provided for the festival.

There were a number of different ways in which we supported the filmmakers.
Firstly at the level of individual films we financed the submission
fees for the top four highest scoring films (`Astroturf', `Dark Matter(s)',
`Murmurs of a Macrocosm', `Noise') to existing UK film festivals,
as this was flagged by the filmmakers as a limiting factor in their
ability to share the work more widely. We left it up to the filmmakers
to determine which festivals might be the best fit for them given
the budget offered to each. Secondly, we acted as champions representing
the entire set of shorts, approached numerous film exhibitors to enquire
about some of them being considered for screening within their existing
events. Finally, we wanted to offer the entire set of shorts as a
ready-made package that could be screened elsewhere. However, it was
deemed that simply showing the shorts back-to-back and not also having
some background to their context and production would not be such
an informative nor entertaining experience for audiences. Therefore,
it was decided to produce a framing film which would incorporate the
shorts into an anthology.

\begin{figure*}
\begin{centering}
\includegraphics[width=0.75\columnwidth]{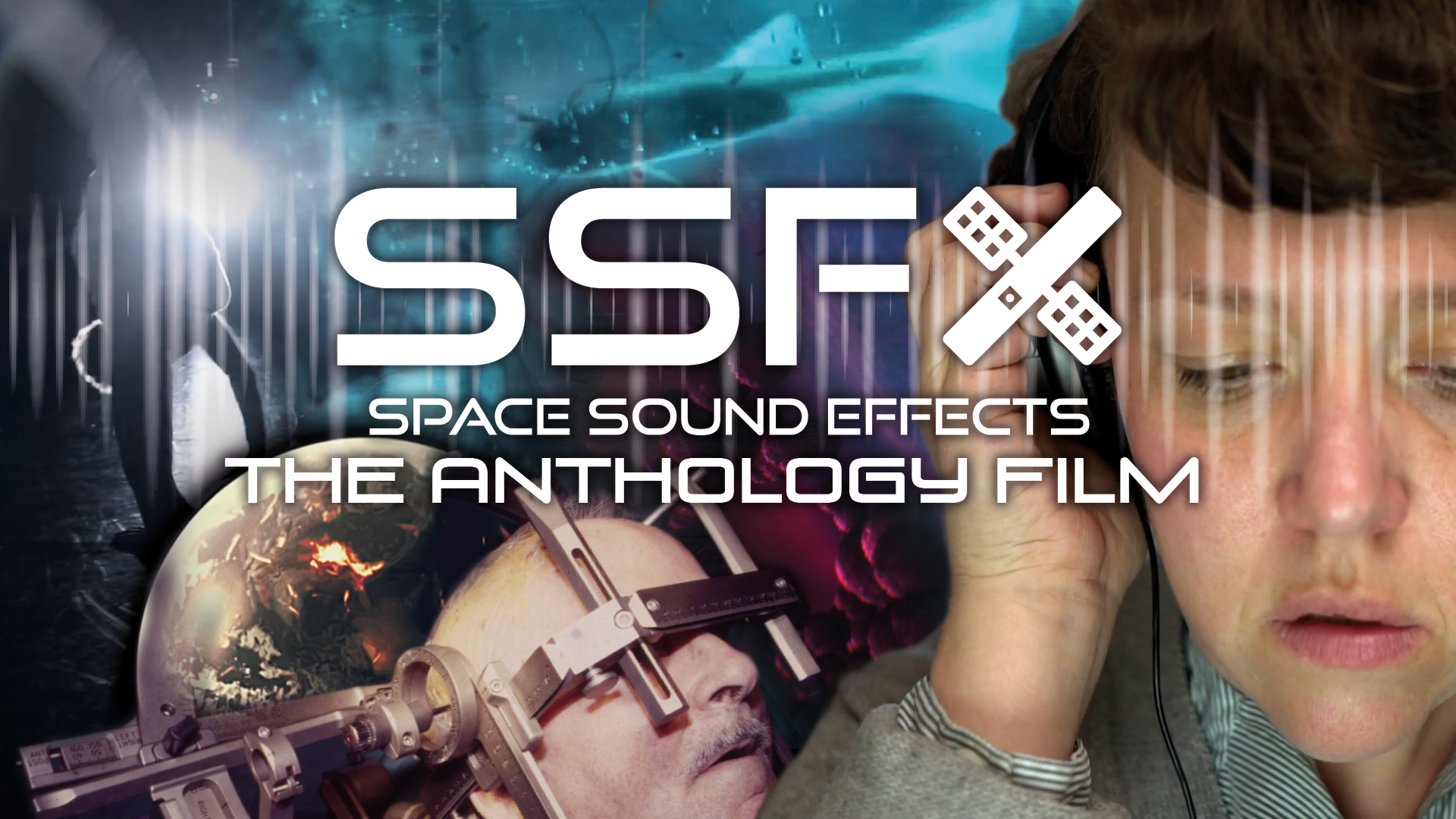}
\par\end{centering}
\caption{Anthology film poster composed of imagery from each of the selected
shorts.\label{fig:Anthology-film-poster}}
\end{figure*}

A tender was put out to various production houses and through networks
soliciting pitches for this framing film. The aim was that this framing
film could in a narrative way bridge together the short films while
also communicating a few simple messages about the underlying science
behind the space sounds and how they affect us. We received three
proposals, which were very different in approach. We went ahead with
one which envisioned a point-of-view shot film during a space weather
event, where technology failing causes the films to appear on various
screens around the house. We worked closely with the production company
on the development of the script and through the production process.
The poster for the anthology film, composed of imagery from the individual
short films, is shown in Figure~\ref{fig:Anthology-film-poster}.
This anthology was released for free on YouTube in October 2018 (https://www.youtube.com/watch?v=P5\_OljSnA1k).

Overall, the filmmakers were very grateful for all the support offered.
One of the filmmakers noted:\begin{quote}\textit{\textquotedblleft The
SSFX project was incredibly rewarding and allowed us as a creative
team to learn about an exciting area we had no knowledge of previously.
We also met other interesting filmmakers and found the audiences incredibly
engaged and interested into the background of the project and where
the research is. We have had a lot of positive feedback and have been
able to direct our audiences to the SSFX website for more information.
Martin Archer has been incredibly supportive and a champion for these
films, for which we are incredibly grateful and have found invaluable.\textquotedblright{}}
(producer of `Astroturf')\end{quote}Another said:\begin{quote} \textit{``Collaborating
on the project was a wonderful experience and we were so grateful
for all the opportunities offered to our film from taking part, reaching
an international audience with our film and getting to enter into
dialogues with audiences, scientists and other filmmakers. If it wasn\textquoteright t
for SSFX, I\textquoteright m not sure that our production team would
have thought to engage so thoughtfully with sounds from space --
we didn\textquoteright t know they existed. Not only did Martin provide
exceptionally unique and compelling sounds for any sound designer
to work with, but he was so thoughtful in terms of explaining the
science and techniques used to capture these sounds. As someone who
was a little intimidated by science in school, I really felt an understanding
around the basic mechanisms and significance around obtaining these
compelling sounds. As a filmmaker, I appreciated this so that I could
answer questions at Q and As with confidence and ease. One of the
other aspects about this project that I appreciate was Martin\textquoteright s
ability to have people from all over the world join in on the conversation.
Multiple times throughout this process, I was able to talk to audience
members and fellow filmmakers on a different continent while staying
in the US. This is not something that I\textquoteright ve experienced
a lot in the independent film festival world. If you can\textquoteright t
physically attend, then you simply can\textquoteright t be a part
of the conversation. That was not the case with SSFX and it made the
experience all the more educational, inclusive and fun for everyone
involved.''} (director of `Dark Matter(s)')\end{quote}These comments
highlight the impact that the open, collaborative and supportive approach
that SSFX took had on the filmmakers.

\section{Infiltrating film events\label{sec:events}}

Several complementary approaches were taken to get the SSFX films
more widely seen as part of film festival and events programmes. As
noted earlier, we paid the submissions fees for film festivals (limited
to the UK only due to funding usage restrictions) identified by the
filmmakers. In addition to this, we advertised the free shorts and/or
anthology through film exhibitor networks recommended by our judges
(including the various BFI Film Audience Network hubs and the Independent
Cinema Office) and approached several key film exhibitors, enquiring
about either arranging one-off screenings of the anthology or showing
the shorts before their features. This generally fed into their aims
of developing audiences for and increasing access to a diversity of
film content for local independent film-going communities. We also
liaised with a few science focused events such as science festivals,
either through open calls or those that approached us, about integrating
the films into their programme in some way.

\begin{sidewaystable*}
\noindent\resizebox{\textwidth}{!}{

\begin{tabular}{cllcccccc}
 & \textbf{Event} & \textbf{Location} & \textbf{Type} & \textbf{Pre-existed} & \textbf{Facilitator} & \textbf{Shown} & \textbf{\# Screenings} & \textbf{Audience}\tabularnewline
\hline 
a & Academia Film Olomouc & Olomouc, Czech Republic & Art-Science & Yes & Filmmaker & Shorts (\#3) & 1 & 70\tabularnewline
b & Aesthetica Short Film Festival & York, UK & Art & Yes & Filmmaker & Shorts (\#3) & 2 & 71\tabularnewline
c & AM Egypt Film Festival & Cairo, Egypt & Art & Yes & Filmmaker & Shorts (\#2) & 1 & 100{*}\tabularnewline
d & British Science Festival & Hull, UK & Science & Yes & Scientist & Anthology & 1 & 25\tabularnewline
e & Cambridge Film Festival & Cambridge, UK & Art & Yes & Filmmaker & Shorts (\#1) & 1 & 100{*}\tabularnewline
f & Central Film School & London, UK & Art-Science & No & Scientist & Anthology & 1 & 10\tabularnewline
g & Escape Velocity & Maryland, USA & Art & Yes & Filmmaker & Shorts (\#3, 5) & 2 & 100{*}\tabularnewline
h & Genesis Cinema & London, UK & Art-Science & No & Scientist & Anthology & 1 & 25\tabularnewline
i & Grand Concourse Film Screning Series & New York, USA & Art & Yes & Filmmaker & Shorts (\#2) & 1 & 100{*}\tabularnewline
j & Imagine Science Film Festival & New York, USA & Art-Science & Yes & Filmmaker & Shorts (\#1) & 1 & 51\tabularnewline
k & IN|DUST|REAL Video Art & Oradea, Romania & Art & Yes & Filmmaker & Shorts (\#2) & 1 & 100{*}\tabularnewline
l & Les Films de la Toile & Paris, France & Art & Yes & Filmmaker & Shorts (\#2) & 1 & 100\tabularnewline
m & Liverpool Film Festival & Liverpool, UK & Art & Yes & Filmmaker & Shorts (\#3) & 1 & 100{*}\tabularnewline
n & London Short Film Festival & London, UK & Art & Yes & Filmmaker & Shorts (\#1) & 1 & 75\tabularnewline
o & Nightstar Cinema & London \& Dorset, UK & Art & Yes & Scientist & Shorts (\#1, 2, 3, 7) & 4 & 333\tabularnewline
p & Nozstock: The Hidden Valley Festival & Hereford, UK & Art & Yes & Scientist & Anthology & 1 & 25\tabularnewline
q & On The Other Side From You, East End Film Festival & London, UK & Art & Yes & Filmmaker & Shorts (\#3) & 1 & 102\tabularnewline
r & Rooftop Film Club & London, UK & Art & Yes & Scientist & Shorts (\#3) & 1 & 155\tabularnewline
s & Royal Society Summer Science Exhibition & London, UK & Science & Yes & Scientist & Shorts (\#1, 2, 3, 5, 7) & 5 & 70\tabularnewline
t & SCInema International Science Film Festival & Various, Australia & Art-Science & Yes & Filmmaker & Shorts (\#1) & 6 & 1,800\tabularnewline
u & SCInema Community Screenings & Various, Australia & Art-Science & Yes & Filmmaker & Shorts (\#1) & 818 & 89,000\tabularnewline
v & Shorts on Tap & London, UK & Art & Yes & Filmmaker & Shorts (\#1) & 1 & 120\tabularnewline
w & SMASHfest & London, UK & Science & Yes & Scientist & Shorts (\#1, 2, 3, 4, 7) & Exhibition & 2,076\tabularnewline
x & South London Shorts & London, UK & Art & Yes & Scientist & Shorts (\#3, 7) & 2 & 80{*}\tabularnewline
y & Southampton International Film Festival & Southampton, UK & Art & Yes & Filmmaker & Shorts (\#1) & 1 & 15\tabularnewline
z & Space - Music \& Film inspired by the cosmos & London, UK & Art-Science & Yes & Scientist & Shorts (\#1) & 1 & 40\tabularnewline
aa & Space Lates & Leicester, UK & Science & Yes & Scientist & Shorts (\#1, 2, 3, 7) & 4 & 200\tabularnewline
ab & SSFX Short Film Festival & London, UK & Art-Science & No & Scientist & Shorts (\#1, 2, 3, 4, 5, 6, 7) & 7 & 85\tabularnewline
ac & SSFX Anthology Premiere & London, UK & Art-Science & No & Scientist & Anthology & 1 & 35\tabularnewline
ad & Storyhouse & Chester, UK & Art-Science & No & Scientist & Anthology & 1 & 10\tabularnewline
ae & Viten Film Festival & Bergen, Norway & Art-Science & Yes & Filmmaker & Shorts (\#6) & 1 & 40\tabularnewline
\end{tabular}

}\caption{List of SSFX events. Individual shorts are numbered as per section~\ref{subsec:submissions}.
Asterisks ({*}) denote estimated audience figures due to lack of information
from event organisers.\label{tab:events}}
\end{sidewaystable*}

Table~\ref{tab:events} details all the events which featured SSFX
film screenings, where these have been grouped by initiative since
in several cases multiple screenings of the same or different films
occurred. There was a large overrepresentation of UK-based events
($68\pm10\%$) compared to all film festivals globally ($8\%$ , $p=3.5\times10^{-17}$)
as listed in \citet{filmfreeway}. This was in part due to funding
usage restrictions limiting which festivals could be applied for,
however, we also note that many film festivals aim to highlight the
works of filmmakers from their own country.

\begin{figure*}
\begin{centering}
\includegraphics{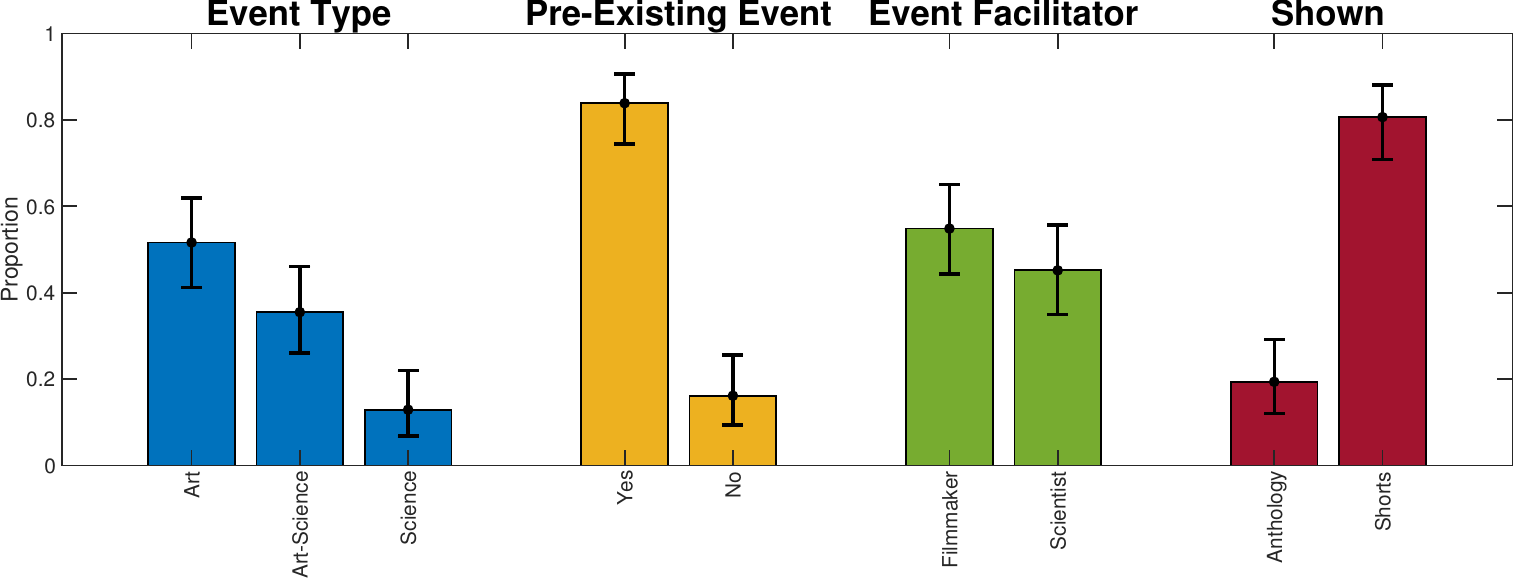}
\par\end{centering}
\caption{Summary statistics of SSFX events.\label{fig:event-stats}}
\end{figure*}

Events have been classified as either art, art-science, or science,
with the distribution of event types shown in Figure~\ref{fig:event-stats}.
Art events denote those with no clear association with science whatsoever,
science events indicate those with no explicit link to art, and art-science
is used to describe events with a stated connection between the two
subjects. There were substantially more art events than science ones
($p=0.012$, $\alpha_{Bonf}=0.017$), which constituted the only significant
difference between SSFX event types. Excluding science events, art-science
($41\pm11\%$) was overrepresented compared to all film festivals
globally that contain some mention of space or science ($11\%$, $p=6.1\times10^{-5}$).
Both of these results reflect some of the struggles faced in the second
phase of the SSFX project --- science event programmers were largely
uninterested in art-science since their audiences are already highly
engaged with science and not necessarily with art, whereas many film
event programmers we approached struggled to understand the concept
of the project thinking the films were aimed at science audiences
rather being open to judge them as films in their own right that happen
to contain a scientific connection (i.e. their preconception was Type
VI of \citet{malina10} rather than Type V).

Figure~\ref{fig:event-stats} demonstrates that screenings predominantly
occurred as part of pre-existing events rather than at bespoke ones
($p=1.9\times10^{-4}$), indicating SSFX was largely successful at
infiltrating science into the film world, and there was a fairly even
split in events arranged by filmmakers or scientists. We note that
filmmakers were more successful at infiltrating art events ($71\pm14\%$
of all their events) than the scientist ($29\pm15\%$), though this
difference was not strictly statistically significant when accounting
for multiple comparisons ($p=0.021$, $\alpha_{Bonf}=0.017$). Primarily
it was an individual short film or subset of the collection of shorts
which was exhibited at events rather than the full anthology film
($p=8.8\times10^{-4}$), which we struggled to convince film programmers
to incorporate into events despite advice from film industry collaborators
that this might be an attractive proposition. Of the individual shorts
`Astroturf' was the most successful, though the only statistically
significant differences ($\alpha_{Bonf}=0.0024$) in the number of
distinct events/initiatives by film were between `Astroturf' and both
`Names and Numbers' ($p=5.0\times10^{-4}$) and `Saturation' ($p=1.9\times10^{-4}$).
We note that neither of these latter two films' festival submission
fees were funded by the project and in the case of `Saturation' a
number of exhibitors expressed that they could not screen it at their
family-friendly events due to the potentially upsetting medical imagery
(edited clips from `Noise' removing the strong language and drug usage
were however able to be used). In terms of total number of screenings,
`Astroturf' had significantly more than all the other shorts though
this was purely due to being selected for the touring SCInema International
Science Film Festival (and associated community screenings) across
Australia.

Film festival acceptance rates are typically $\sim5\%$ \citep{stephenfollows13b}
with the largest festivals being $\lesssim1\%$ \citep{festivalwhizz}.
While we do not have concrete numbers on exactly how many festivals
the filmmakers submitted to, given the budget and average submission
cost for short films \citep{stephenfollows13a} we estimate around
30 total submissions. This means that the 17 festival successes constitutes
an impressive acceptance rate across the shorts of $57\pm11\%$, significantly
higher than expected. This perhaps reflects the quality of the art-science
films that resulted from the project. We also note that given the
filmmakers were submitting their shorts to festivals independently
and all found success, this lends confidence beyond just an individual
case study that this model of infiltrating science into cultural events
can indeed work.

\section{Engaging audiences through film\label{sec:audiences}}

We generally relied on the event organisers to attract audiences,
since they have built-in audience bases from their previous activities.
Given we were largely infiltrating existing events, this limited the
evaluations that could be implemented especially as at many events
(especially the international ones) no filmmakers or scientists from
the project were physically present. Therefore, evaluation data was
typically collected only at bespoke SSFX events and several methods
were employed: a ball and bin question upon arrival assessing prior
knowledge, graffiti walls at post-film receptions assessing their
motivations and takeaways, and an online survey three weeks later
for those who left contact details. Filming by a third party at the
SSFX Short Film Festival (event ab) captured additional qualitative
data. Given that these events where evaluation was possible tended
to show all the shorts (either individually or via the anthology)
we are unable to comment on whether certain SSFX films were more impactful
upon attendees than others.

\begin{figure}
\centering{}\includegraphics{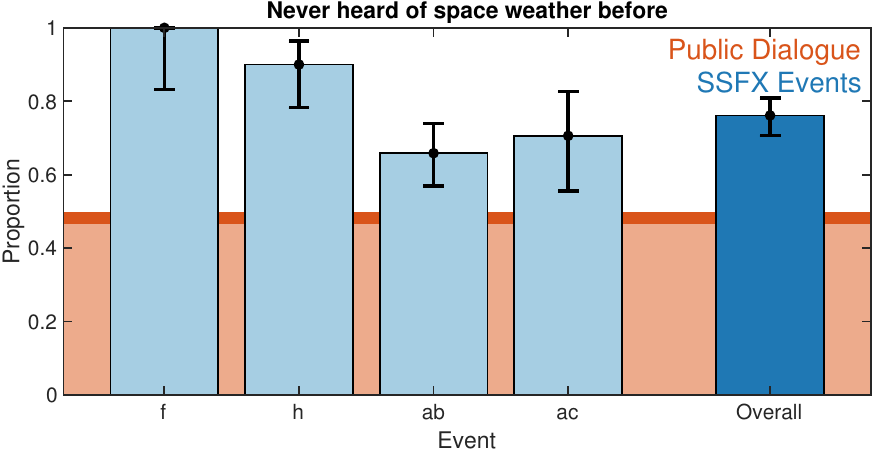}\caption{Prior knowlege of audience at SSFX art-science events in blue compared
to a recent public dialogue \citep{spaceweatherdialogue15} in orange.\label{fig:knowledge}}
\end{figure}

As part of a recent public dialogue, \citet{spaceweatherdialogue15}
found in a survey of 1,010 people representative of the UK adult population
(by gender, age, social grade, education, dependants, geographic region,
and human settlement type) that $48\pm2\%$ have never heard of space
weather before. We therefore asked audiences upon arrival at events
whether they had heard of space weather before, via a ball and bin
method where attendees were instructed to put a ball in either the
`yes' or `no' bin. The results from the individual events where we
asked this question are displayed as the light blue bars in Figure~\ref{fig:knowledge}
indicating levels greater than in the public dialogue (orange). Combining
the data from all these SSFX events gives an overall result (dark
blue) of $76\pm5\%$, which constitutes $2.95\pm0.20$ times more
likely (the odds ratio) to have never heard of space weather than
the general population ($p=9.2\times10^{-8}$). Therefore an atypical
audience was attracted to these events in terms of prior knowledge.
Note that these results came exclusively from art-science events and
arguably one might expect an even greater overall proportion of people
to be unaware of the field at the art events that SSFX infiltrated.

\begin{figure}
\begin{centering}
\includegraphics{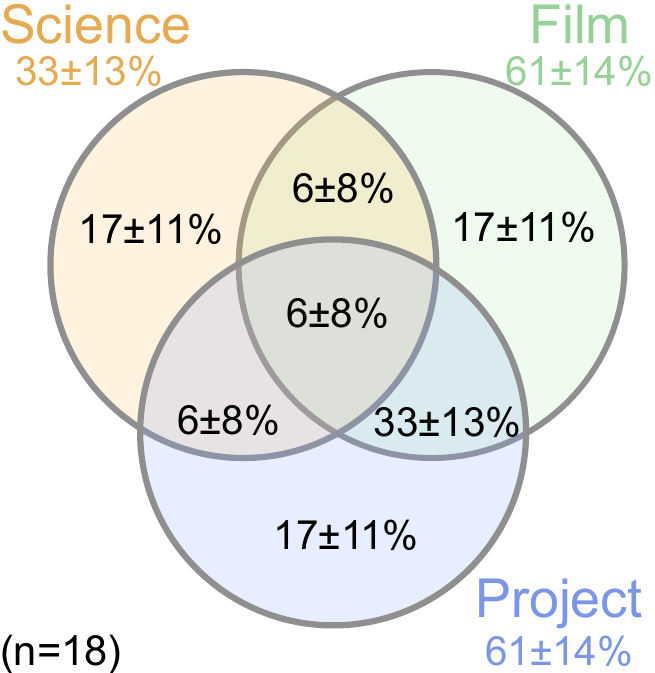}
\par\end{centering}
\caption{Venn diagram of people's motivations for attending SSFX art-science
events in the same format as Figure~\ref{fig:filmmaker-venn}.\label{fig:audience-venn}}

\end{figure}

Another way we assessed whether the project attracted new audiences
was by asking what motivated them to attend. At two art-science events
(ab and ac) this was collected via open-ended graffiti walls, where
9 responses were recorded which can be found in Appendix~\ref{sec:Qualitative-data}.
Through thematic analysis \citep{braun06} it was possible to group
all of these as being due to an interest in science (e.g.\textit{
``love science''}), film (e.g. \textit{``I like weird films''}),
or specifically the project (e.g. \textit{``interesting concept''}),
where the quotes displayed serve as representative illustrative examples
from different respondents. Follow-up online surveys after several
events (f, h, aa, ab, and ac) specifically asked in a closed question
whether attendees had been attracted due to regular attendance at
science events, film events, or if it was specifically this event
that had interested them. They could select as many options as were
applicable. Given this yielded only 12 responses we opt to combine
the data from both methods, omitting event aa since out of those events
surveyed it was the only science event as well as the only pre-existing
one. The overall results are shown in Figure~\ref{fig:audience-venn}.
Repeating the same analysis as with filmmakers' motivations revealed
that significantly more people attended due to being film-goers or
specifically being interested in the project ($F\cup P$) at $78\pm12\%$
compared to attending science events often ($S$) at $33\pm13\%$
($p=0.0023$, $\alpha_{Bonf}=0.0083$). This therefore provides further
evidence that SSFX was able to attract substantial non-science audiences,
placing it as comparable to some of the most successful art-science
events across different art forms at reaching new audiences \citep[cf.][]{brook17}.
Again we note that since this analysis pertained only to art-science
events, it is highly likely at the art events SSFX infiltrated that
even fewer people would have exhibited science interests given the
complete lack of a science-connection at these events and the existing
research into the motivations behind film festival attendence \citep{baez14}.

\begin{figure*}
\begin{centering}
\includegraphics{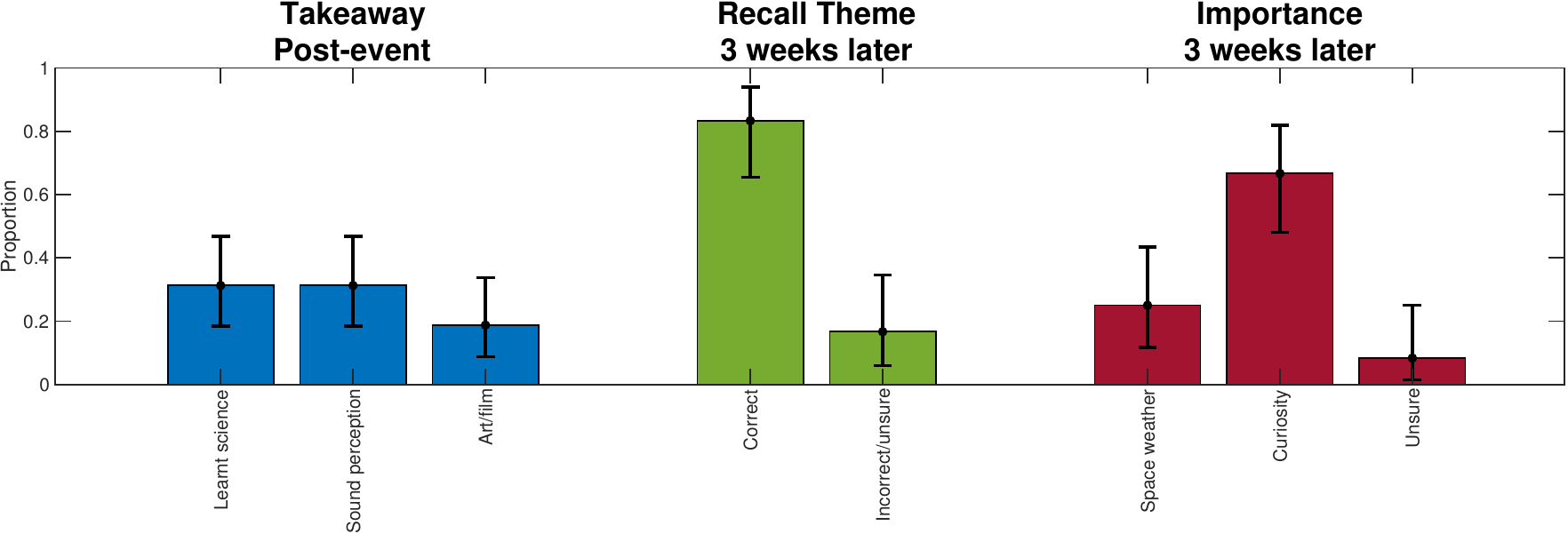}
\par\end{centering}
\caption{Summary statistics of evaluation data following SSFX events.\label{fig:eval-stats}}

\end{figure*}

We assessed the learning outcomes of attendees through the follow-up
surveys, asking in open questions if they recalled the event's theme
and why this topic is studied/important. The full list of responses
is given in Appendix~\ref{sec:Qualitative-data}. As shown in Figure~\ref{fig:eval-stats},
the majority ($83\pm14\%$, $p=0.039$) correctly recalled that the
event concerned the sounds of space or provided more specific answers.
Interestingly most ($67\pm17\%$) thought this area was important
due to the inherent value of science / intellectual curiosity (e.g.
\textit{``It helps us to understand the universe, physics, and gives
us a clearer idea of the world around us''}) rather than citing space
weather ($25\pm16\%$), though this majority was not statistically
significant and neither were the differences between responses. As
far as we are aware there is little published research into the recollection
of public engagement events' themes and key messages by attendees
in follow-up surveys. However, comparing with studies into the recollection
of television campaigns \citep[e.g.][]{berry09,potter19} and the
so-called Memory Chain Model \citep{murre15} suggests that the fraction
quoting space weather would be deemed successful, while the recollection
of the event's overall theme would be considered extremely high. Also,
given the atypical non-science audience, the fact that many attendees
took away from the event the value of fundamental scientific research
was an unanticipated but very welcome impact.

In terms of impact on attendees, at two art-science events (ab and
ac) we asked via graffiti wall what (if anything) they had gained
or taken away from the event. Most of the 16 responses, displayed
in full in Appendix~\ref{sec:Qualitative-data} with representative
examples from different respondents shown here, could be broadly categorised
using grounded theory \citep{silverman10,robson11} as concerning
the science (e.g. \textit{``amazing space sounds I want to learn
more about''}), perception of sound (e.g. \textit{``people hear
differently''}), or art/film (e.g. \textit{``grown an interest in
film-making''}), with the proportions of each shown in Figure~\ref{fig:eval-stats}
demonstrating a near even split between the three. Other miscellaneous
takeaways included aspects of science communication, humanity, and
specific (non-science) themes raised in the films. Furthermore, at
the SSFX Short Film Festival (event ab) a selection of people were
interviewed during the reception by a third party, with some of the
responses available at https://www.youtube.com/watch?v=sPa7avaksFI.
The most common point that emerged (again analysed using grounded
theory) was that attendees really enjoyed the broad range of interpretations
of the same space sounds which were expressed in the different films.
Others commented on how the concept of the festival was an interesting
approach of bringing scientific ideas to a wider audience, that they
had learned about and gained an interest in the science behind the
sounds, and that it attracted a diverse group of people with a lot
of interaction particularly in the reception. On this latter point,
it was anecdotally noted at most of the events that the diversity
of audiences by gender and ethnicity appeared much greater than compared
to typical physics engagement events, though this was not captured
quantitatively. Respondents in the follow-up survey also noted other
takeaways from attending which we again have coded using grounded
theory and provide illustrative example responses from different participants
here: enjoying or being inspired by the event (e.g. \textit{``Really
enjoyed the enthusiasm of the speaker and the topic of the films mixed
with science''}), the creativity/diversity of films (e.g. \textit{``how
each filmmaker found the humanity in sounds from space''}), meeting
and hearing from both scientists and filmmakers (e.g. \textit{``It
was interesting to meet some of the people involved in both the science
and filmmaking''}), and the importance/relevance of the scientific
research (e.g. \textit{``Genuine and relevant science research and
knowledge is vital and underused in the film industry''}). Of the
responses which did not fit into any of these themes, one person said
that they had developed an interest in arts events by attending (\textit{``I
will definitely look at the }{[}arts venue{]}\textit{ Rich Mix website
more for future events''}) while another found the anthology film
to be \textit{``very strange''}, which may or may not be a good
thing. One respondent wrote in detail on their thoughts of the virtues
of this type of art-science collaboration:\begin{quote}\textit{\textquotedblleft Taking
raw data out of context and using it as a key creative element in
the creation of art is a way of providing a fresh look at a scientific
inquiry. Art can be a mirror whose reflection can reset context and
provide the listener with a different perspective than might otherwise
be encountered. The result of this competition has been a number of
submissions that stimulate a wider audience to think about how science
is more than just the collection of raw data, and that understanding
can come from looking at results from a new vantage.\textquotedblright }\end{quote}We
note that despite the somewhat limited evaluation data, it does not
appear that the impacts from events which exhibited the short films
(with their prologue and epilogue text concerning the science) are
significantly different from those of the anthology film (which contained
substantial additional messaging through the bridging film). The overall
results highlight that there were many unforeseen impacts upon attendees
outside of simply raising awareness of the research area to atypical
audiences.

\conclusions{}

The SSFX (Space Sound Effects) Short Film Festival was an art-science
collaboration project aimed at infiltrating space science into culture
through the medium of film. In particular it invited the usage of
sonified satellite data of plasma waves in Earth's magnetosphere,
a key component within space weather, as key creative elements.

The first audience the project aimed to engage were independent filmmakers
through challenging them to use these space sounds to create short
films. Through partnership with film industry experts and organisations,
an international film festival was run adopting many of the standard
practises within the sector to lend authenticity and legitimacy to
the project. Formative evaluation of people who registered interest
with the project during the submission revealed that we successfully
hooked the filmmaking community, though most who engaged also had
a general interest in science. Seven very different films were selected
for screening. Feedback from these filmmakers highlighted that they
relished the creative freedom afforded to them in interpreting the
sounds and their usage within their works, hence very open criteria
are not only enticing to filmmakers but also enable a broad range
of art works to be produced. Another important aspect to the project
was in supporting the filmmakers and championing their films after
the initial festival, which had the mutual benefit of raising the
profile of the filmmakers whilst also sharing the underlying science
more widely.

The second audience was film programmers and exhibitors in trying
to infiltrate the produced short films into existing events. While
an anthology film packaging all the shorts together through a science-based
narrative was produced, we struggled to get this shown and found much
greater success with the individual short films. Filmmakers were best
placed to submit their own works to film festivals following the standard
method, with monetary support from the scientists, as they have a
better idea of which festivals would be most appropriate. However,
scientists were still able to play a role in representing the full
suite of shorts for consideration at other sorts of film events. Both
of these approaches led to SSFX infiltrating more art events than
science ones, as desired, though a substantial number of art-science
events also occurred.

The project ultimately also aimed to raise awareness of the science
to atypical audiences through the use of the films. While audience
evaluation proved challenging due to SSFX films typically sitting
within larger events organised by others, some evaluation was able
to be done at mostly bespoke art-science events. This highlighted
that attendees were much less aware of the topic of space weather
than the general public and were much more likely to have attended
due to an existing interest in film or specifically the concept of
SSFX rather than having an existing science interest. This placed
the project as comparable to some of the most successful art-science
events across different art forms at reaching new audiences. Many
different, and often unanticipated, impacts were had on attendees
beyond simply learning about the science, which demonstrates the versatility
of film as a form of art at provoking varied responses in audiences.

We therefore advocate that adopting a film festival model can result
in creative art-science that fits within the many film-based cultural
events around the world. This enables the power of cinema to be leveraged
on audiences that don't normally engage with science, thus providing
one potential means of breaking beyond the scientific ``echo chamber''
in perveying the importance and relevance of scientific research.

\appendix

\section{Statistical techniques\label{sec:Statistical-techniques}}

Several statistical methods are used throughout this paper which are
detailed here.

All uncertainties quoted or displayed, e.g. through errorbars, represent
standard (i.e. 68\%) intervals. For proportions/probabilities these
are determined through the \citet{clopper34} method, a conservative
estimate based on the exact expression for the binomial distribution,
and therefore represent the expected variance due to counting statistics
only and not any other potential sources.

Several statistical hypothesis tests are used with effect sizes and
two-tailed $p$-values being quoted. Throughout the desired significance
level $\alpha$ is set as $0.05$, though in the case of multiple
comparisons we use the Bonferonni correction where the significance
level per comparison is $\alpha_{Bonf}=\alpha/N$ for $N$ total possible
comparisons. Two-tailed binomial tests are used to compare proportions
of both independent and correlated (i.e. within the same) samples.

Finally, the agreement between judges scores is quantified using the
alpha coefficient of \citet{krippendorff70,krippendorff}, which is
computed as unity minus the ratio of the observed disagreement to
that expected by chance, i.e.
\[
\alpha=1-\frac{\mathop{\frac{1}{n}\sum_{c}\sum_{k}}o_{ck}\delta_{ck}^{2}}{\frac{1}{n\left(n-1\right)}\sum_{c}\sum_{k}n_{c}n_{k}\delta_{ck}^{2}}
\]
where $o_{ck}$ are the observed frequencies in a coincidence matrix,
$n_{c}$ are the column totals in this matrix, $n$ is sum of the
entire matrix, and $\delta_{ck}$ is a metric function for which we
use the one applicable to ordinal data. The intepretation of this
coefficient is that a value of 1 indicates perfect agreement between
judges, 0 would result from randomly drawn scores, and a negative
value is possible when disagreements are systematic and exceed what
can be expected by chance.

\section{Qualitative data\label{sec:Qualitative-data}}

Here we tabulate the various qualitative data captured from audiences
at events, where each row contains responses from a single unique
participant. The qualitative data was coded and analysed by the author
using thematic analysis \citep{braun06}, however, no a priori codes
were generated instead allowing these to naturally emerge from the
data via a grounded theory approach \citep{silverman10,robson11}.
The final themes determined by this method and their association to
the raw qualitative data are also listed in the following tables.

Firstly, the responses from researchers and public engagement professionals
to open questions through an interactive online survey during an art-science
session at the 2019 Interact symposium \citep{interactsession19}
were as follows.\begin{footnotesize}

\begin{longtable}[c]{>{\raggedright}p{0.3\textwidth}>{\raggedright}p{0.3\textwidth}>{\raggedright}p{0.3\textwidth}}
\textbf{Why are you interested in art-science collaborations?} & \textbf{Who do you think are the target audience in art-science?} & \textbf{Why do you think art-science is important?}\tabularnewline
\hline 
I want to build creativity into my role & A wide range of audiences, beyond academia & Both pursuits are creative - potential to communicate in new ways\tabularnewline
{[}Blank{]} & Anyone interested in art. & Different ways of communicating science\tabularnewline
To communicate abstract concepts & Everyone & Increase fascination and curiosity in BOTH subjects\tabularnewline
Worked well in the past. Would like to find out more! & People who do not normally engage with science & {[}Blank{]}\tabularnewline
I enjoy both and don't see why I should have to choose between them. & People who think they like art but not science & Science should be an embedded part of culture\tabularnewline
It\textquoteright s important to combine them together & Everyone should be as science and art is everywhere & It\textquoteright s everywhere whether you realise it or not\tabularnewline
I'm a scientist and my husband is an artist. We collaborate and I'm
interested to see how others combine the subjects. & Everyone & Communication tool\tabularnewline
I work with communities that do not have English as their first language
and art crosses language barriers & Anyone and everyone & Reaches new audiences and can illustrate the science in new exciting
ways\tabularnewline
Provides new creative accessible ways to access science and vice versa & Depends on purpose of engagement activity & Need to enable conversations and accessibility and tap into interests
/ challenges\tabularnewline
It is fun and very creative! & Anyone! Crosses all spectrums & To get a science message across in a unique thought provoking way
and challenge artists to enage in new areas\tabularnewline
To describe science in interesting new ways & Everyone & Because it can explain quite abstract concepts in a unique way\tabularnewline
Engage with new audience & Everyone & Art and science are both creative both can learn from each other.\tabularnewline
\end{longtable}

\end{footnotesize}\bigskip{}
The following table displays the responses from audiences captured
on a graffiti wall concerning what attracted them to SSFX art-science
events.

\begin{footnotesize}

\begin{longtable}[c]{clccc}
\multirow{2}{*}{\textbf{Event}} & \multirow{2}{*}{\textbf{What attracted you to this event?}} & \multicolumn{3}{c}{\textbf{Themes}}\tabularnewline
 &  & Science & Film & Project\tabularnewline
\hline 
ab & Collab-lab {[}factual science filmmakers{]} tweet + interest in subject & \checkmark &  & \tabularnewline
ab & Interesting concept &  &  & \checkmark\tabularnewline
ab & Email on MIST {[}Magnetosphere Ionosphere Solar Terrestrial community{]} & \checkmark &  & \tabularnewline
ac & Short films with an awesome original soundtrack &  & \checkmark & \checkmark\tabularnewline
ac & Interested in the brief to create films from such an unusual sound
source &  & \checkmark & \checkmark\tabularnewline
ac & Space + sound art-science crossover &  &  & \checkmark\tabularnewline
ac & Love science and sound design & \checkmark &  & \checkmark\tabularnewline
ac & I like weird films &  & \checkmark & \tabularnewline
ac & I wanted to see what filmmakers could do with space sounds &  & \checkmark & \checkmark\tabularnewline
\end{longtable}

\end{footnotesize}\bigskip{}
A similar graffiti wall also asked audiences what they felt that they'd
gained from attending these events.

\begin{footnotesize}

\begin{longtable}[c]{c>{\raggedright}p{0.5\textwidth}ccc}
\multirow{2}{*}{\textbf{Event}} & \multirow{2}{0.5\textwidth}{\textbf{What if anything have you gained or taken away from this event?}} & \multicolumn{3}{c}{\textbf{Themes}}\tabularnewline
 &  & Learn science & Sound perception & Art/Film\tabularnewline
\hline 
ab & Ideas on how to explain space weather &  &  & \tabularnewline
ab & National anthem as cultural boundary &  &  & \tabularnewline
ab & Noise &  & \checkmark & \tabularnewline
ab & Space weather & \checkmark &  & \tabularnewline
ab & Use sound to promote my research and get \$\$ &  & \checkmark & \tabularnewline
ab & Chirp &  & \checkmark & \tabularnewline
ab & People hear differently &  & \checkmark & \tabularnewline
ab & Please see ''no'' ball container & \checkmark &  & \tabularnewline
ab & Ideas of being human &  &  & \tabularnewline
ab & Grown an interest in film-making &  &  & \checkmark\tabularnewline
ab & Perceptions of space sounds &  & \checkmark & \tabularnewline
ab & .- .-. - (art) is a mirror &  &  & \checkmark\tabularnewline
ac & Amazing space sounds I want to learn more about it & \checkmark &  & \tabularnewline
ac & I want to learn more about the science behind it & \checkmark &  & \tabularnewline
ac & I had never heard of SSFX before but now want to hear more & \checkmark &  & \tabularnewline
ac & The theme of the narrative linking segments put me in mind of the
Hungarian film `Adas' (Transmission) where all electronics mystreriously
die &  &  & \checkmark\tabularnewline
\end{longtable}

\end{footnotesize}\bigskip{}
Finally, the results from an online survey three weeks following various
SSFX events were as follows.

\begin{tiny}

\begin{landscape}
\begin{longtable}[c]{c>{\raggedright}p{0.1\textwidth}>{\raggedright}p{0.16\textwidth}>{\centering}p{20pt}c>{\raggedright}p{0.12\textwidth}>{\raggedright}p{0.2\textwidth}>{\centering}p{20pt}>{\centering}p{22pt}>{\centering}p{20pt}c}
\multirow{2}{*}{\textbf{Event}} & \multirow{2}{0.1\textwidth}{\textbf{What was the theme of the event?}} & \multirow{2}{0.16\textwidth}{\textbf{Why is this topic important / studied?}} & \multicolumn{2}{c}{\textbf{Themes}} & \multirow{2}{0.12\textwidth}{\textbf{Is there anything else you gained or took away from the event?}} & \multirow{2}{0.2\textwidth}{\textbf{Any comments you'd like to feed back?}} & \multicolumn{4}{c}{\textbf{Themes}}\tabularnewline
 &  &  & Space weather & Curiosity &  &  & Enjoyed / Inspired & Creativity / Diversity & Artists \& Scientists & Science\tabularnewline
\hline 
ab & Space Sounds & Space is bigger than anything human and it gives you a very different
perspective on the human world and the issues that face us. &  & \checkmark & The importance of space and continually exploring the unknown. & Taking raw data out of context and using it as a key creative element
in the creation of art is a way of providing a fresh look at a scientific
inquiry. Art can be a mirror whose reflection can reset context and
provide the listener with a different perspective than might otherwise
be encountered. The result of this competition has been a number of
submissions that stimulate a wider audience to think about how science
is more than just the collection of raw data, and that understanding
can come from looking at results from a new vantage. &  & \checkmark & \checkmark & \checkmark\tabularnewline
ab & Space sounds & It helps us to understand the universe, physics, and gives us a clearer
idea of the world around us &  & \checkmark & I enjoyed the creativity and diversity shown in the films, and how
each filmmaker found the humanity in sounds from space & Very well put together, good programming, enjoyed the presentation
at the start. & \checkmark & \checkmark &  & \checkmark\tabularnewline
ab & Sound in space & It's interesting because the films allow people access to science/physics
in an accessible way and interpret something that happens quite far
away onto a human level. &  & \checkmark & I will definitely look at the Rich Mix website more for future events & The event wasn't sign posted. Made it a bit tricky to know when we're
in the correct place &  &  &  & \tabularnewline
ab & Science communication using sound recordings from space & Facinating but not important... &  & \checkmark & Genuine and relevant science research and knowledge is vital and underused
in the film industry. & This festival needs to grow and expand! It is important! &  &  &  & \checkmark\tabularnewline
aa & Measuring background magnetic variation around earth & For satellites to operate properly and to improve our understanding
of the universe & \checkmark &  & Really enjoyed the creativity of the films! & Very slickly run, might have been nice to intersperse the talk with
the videos rather than all in one. &  & \checkmark &  & \tabularnewline
aa & Unsure & Unsure &  &  & N/A & Enjoyable evening, fascinating talk & \checkmark &  &  & \checkmark\tabularnewline
aa & Sounds from space & Gives us ideas about how the universe works &  & \checkmark & Really enjoyed the enthusiasm of the speaker and the topic of the
films mixed with science & Brilliant. & \checkmark &  & \checkmark & \checkmark\tabularnewline
ac & Sounds from Space & Because Space is cool &  & \checkmark & It was very strange & {[}Blank{]} &  &  &  & \tabularnewline
ac & Life & It was very interesting. &  & \checkmark & Inspiration. & {[}Blank{]} & \checkmark &  &  & \tabularnewline
h & Space, space storms & To see, gauge potential impact/predict/prevent & \checkmark &  & An awareness that this existed & {[}Blank{]} &  &  &  & \checkmark\tabularnewline
h & Fluctations in the magnetosphere & To protect satellites and power grids from electromagnetic interference
- for example a repeat carrington event & \checkmark & 	\checkmark & It was interesting to meet some of the people involved in both the
science and filmmaking & We had fun! & \checkmark &  & \checkmark & \tabularnewline
f & Sound from space used in films & Physics and Films is fascinating &  & \checkmark & I really enjoyed listening to Dr Martin Archer learned a lot & I hope another one can be organised & \checkmark &  & \checkmark & \checkmark\tabularnewline
\end{longtable}
\end{landscape}\end{tiny}

\dataavailability{Data supporting the findings of this study are
contained within the article or derived from listed public domain
resources.}

\authorcontribution{MOA conceived the project and its evaluation,
performed the analysis, and wrote the paper.}

\competinginterests{The author declares that they have no conflict
of interest.}
\begin{acknowledgements}
We thank the SSFX Short Film Festival judges Laura Adams, David Berman,
Ed Prosser, and Jake Roper; the selected filmmakers Adam Azmy, Victor
Galv\~{a}o, Nidhi Gupta, Aaron Howell, Ali Jennings, Simon Rattigan,
Jesseca Ynez Simmons, and James Uren; and all the film industry experts
and exhibitors that helped us share this work with audiences. This
project was supported by a QMUL Centre for Public Engagement Large
Award, EGU Public Engagement Grant, and STFC Public Engagement Spark
Award ST/R001456/1.
\end{acknowledgements}
\bibliographystyle{copernicus}
\bibliography{ssfx}

\end{document}